\documentclass[prl,twocolumn,superscriptaddress,showpacs]{revtex4-1}
\usepackage{amsmath}%
\usepackage{amsfonts}%
\usepackage{amssymb}%
\usepackage{graphicx}
\usepackage{color}
\usepackage{tabularx}
\usepackage{braket}

\usepackage{comment} % comment-out piece of text
\usepackage[normalem]{ulem} %need for strikethrough
\usepackage{cancel} % strike-out in mathmode
\usepackage{color} %needed for coloring of \todo etc.

\begin{document}

% title
\title{Single electron relativistic clock interferometer}

\author{P. A. Bushev}
\affiliation{Experimentalphysik, Universit\"{a}t des Saarlandes, D-66123 Saarbr\"{u}cken, Germany}

\author{J. H. Cole}
\affiliation{Chemical and Quantum Physics, School of Science, RMIT University, Melbourne 3001, Australia}

\author{D. Sholokhov}
\affiliation{Experimentalphysik, Universit\"{a}t des Saarlandes, D-66123 Saarbr\"{u}cken, Germany}

\author{N. Kukharchyk}
\affiliation{Experimentalphysik, Universit\"{a}t des Saarlandes, D-66123 Saarbr\"{u}cken, Germany}

\author{M. Zych}
\affiliation{Centre for Engineered Quantum Systems, School of Mathematics and Physics, The University of Queensland, St Lucia, Queensland 4072, Australia}

\date{\today}

\begin{abstract}
Although time is one of the fundamental notions in physics, it does not have a unique description. In quantum theory time is a parameter ordering the succession of the probability amplitudes of a quantum system, while according to relativity theory each system experiences in general a different proper time, depending on the system's world line, due to time to time dilation. It is therefore of fundamental interest to test the notion of time in the regime where both quantum and relativistic effects play a role, for example, when different amplitudes of a single quantum clock experience different magnitudes of time dilation. Here we propose a realization of such an experiment with a single electron in a Penning trap. The clock can be implemented in the electronic spin precession and its time dilation then depends on the radial (cyclotron) state of the electron. We show that coherent manipulation and detection of the electron can be achieved already with present day technology. A single electron in a Penning trap is a technologically ready platform where the notion of time can be probed in a hitherto untested regime, where it requires a relativistic as well as quantum description.
\end{abstract}


%\keywords{General relativity, Trapping of single electron, Decoherence}

% Make the title.
\maketitle

\section*{Introduction}
One of the most intriguing features of general relativity (GR) is that the time is not absolute but can flow differently for different observers~\cite{Einstein1916}. The physical significance of this essential principle was still widely debated in 1970s~\cite{Terrell1972}.
Later, time dilation effects have been tested in numerous textbook experiments. In 1971, Hafele and Keating synchronized four cesium-beam atomic clocks with a reference clock at the US Naval Observatory and flew them around the world in commercial jets. The agreement between the expected and observed time dilations of $\sim100$~ns confirmed the predictions of both special and general relativity~\cite{Hafele1972}. The most recent time dilation tests performed with atomic clocks have attained the precision ($\sim10^{-18}$) which would allow resolving the gravitational time dilation on the length scale of 2~cm~\cite{Ye2015}. Results of such experiments, including~\cite{Chou2010}, are in agreement with the picture of clocks measuring proper time elapsing along their classical trajectories. In other words, all to date tests of time dilation constitute the realization of the famous twin paradox thought-experiment, in which twins A and B leave from one spacetime event along separate paths, and when they meet again, they discover that they have aged differently -- since proper time elapsing along different world lines is in general different according to relativity theory~\cite{Fock1964}.

Quantum mechanics predicts that any system can propagate along different spacetime paths in a superposition. This has been verified in numerous experiments starting from electron diffraction to atoms~\cite{Andrews1997} and complex molecules~\cite{Eibenberger2013}. Therefore, it would be of fundamental interest to test how relativistic time dilation applies in the case of single particle (clock) propagating along different world lines in quantum superposition. Such an experiment would represent a quantum version of the twin paradox, where a ``single'' twin travels in superposition along different paths. The realization of such a quantum twin paradox experiment would reveal whether a single clock can be prepared in a superposition of different proper times and verify new effects predicted in refs.~\cite{Zych2011,Zych2012,Pikovski2015}. Furthermore, one can ask what would be observed in an experiment with entangled twins, i.e.~where the world lines of two clocks are entangled. As a consequence of both quantum theory and relativity, the clocks should become entangled in their proper times. In principle, one could realize the Bell test with such entangled clocks~\cite{Zych2012}. A violation of the Bell inequality in such a test would demonstrate that proper time cannot be consistently described by any local-realistic variable if both quantum theory and relativity are valid. More generally, such Bell and interference experiments would be testing different sets of hidden variable theories~\cite{Belinfante2014survey} for time.

In addition to fundamental insights into the joint foundations of quantum and relativistic physics, the understanding of the influence of proper time on quantum systems can also contribute to the research on relativistic quantum information. This active research field aims at exploring how special and general relativity will affect our future quantum communication and computation technology~\cite{Mann2012}.

In this article, we propose an experimental implementation of the quantum twin paradox with a single electron quantum cyclotron. The described experimental setup is also promising for the Bell tests of proper time. After a short review of the clock interferometry, we introduce the single electron quantum cyclotron and describe a technique which allows for the coherent control of its cyclotron and spin degrees of freedom. Our feasibility study reveals that the proposed setup can be realized with existing technology.

\section*{Clock interferometer}
%\tosub{Gravitational effects in quantum mechanics were first tested using neutron interferometry~\cite{Coleila1975}. The phase shift measured in this experiment was a consequence of a different gravitational potentials at two different heights. Such interferometric techniques are now used for high precision measurements of gravitational acceleration~\cite{Peters1999}. In recent experiments, the quantization of energy levels of neutrons trapped in a gravitational potential well was observed~\cite{Nesvizhevsky2002,Jenke2011}. However, all quantum experiments performed to date are still fully compatible with the non-relativistic, Newtonian gravity.}

The idea of clock interferometry \cite{Zych2011} was motivated by the search for experiments which would test relativistic gravity effects on quantum systems (and would thus go beyond Newtonian gravity effects, like the gravitational phase-shifts in neutron~\cite{Coleila1975} or atom~\cite{Peters1999} interferometry, or quantization of the energy levels of  neutrons trapped in a gravitational potential~\cite{Nesvizhevsky2002,Jenke2011}.)
Most generally, however,  clock interference experiments can test the effects of the relativistic proper time on quantum coherence. The key aspect of this approach is to use in interferometry a quantum particle that has some internal dynamical degrees of freedom -- a clock. One then considers a scenario where the clock takes in superposition different paths, prepared such that different proper time elapses along each of them. If the relativistic time dilation applies to a clock in a superposition, the internal state of the clock should become correlated with the path. Thus, in the relativistic clock interferometer time dilation is predicted to yield information about the path of the clock, so-called ``which-way'' information. According to the principle of quantum complementarity, the visibility of the interference pattern must therefore decrease~\cite{Englert1996}, to the extent to which the final clock states could allow determining which path has been taken. The interference disappears when the proper time difference is equal to half the clock period: when the internal degrees of freedom propagating along different paths evolve into orthogonal quantum states, and therefore carry maximal which-way information. For a clock with period $T_{clock}$ and the proper time difference $\Delta\tau$ between the superposed paths, the visibility simply reads $V=\vert\cos\pi\Delta\tau/T_{clock}\vert$, see~\cite{Zych2011}. In the non-relativistic case, where time is absolute, the visibility in principle stays maximal, regardless of the paths of the clock. Therefore, observation of such visibility modulation would directly test the effects of relativistic proper time on quantum coherence. Moreover, this visibility modulations would be a strong evidence of the time dilation induced decoherence, predicted to effectively limit quantum coherence for highly complex, macroscopic particles~\cite{Pikovski2015}. Such an effect could have relevance for understanding of how the classical physical laws emerge from the laws of quantum mechanics~\cite{SchlosshauerBook2007}.

In order to probe the reduction of the visibility, relativistic time dilation has to be commensurable with the `ticking' period of the clock. For this reason, direct probing of time dilation caused by gravity is very challenging. Vertical separation of atomic clocks operating at the frequency of 10$^{15}$~Hz should be $\sim$10~m and be maintained in superposition for the time exceeding 1~sec. An interference experiment, where time dilation has been simulated by introducing frequency detuning for clocks travelling along different paths has been recently demonstrated with rubidium atoms in a Bose-Einstein-Condensate~\cite{Margalit2015}.  The clock rate in this experiment was not high enough to probe effects attributed to special or general relativity. Even the state-of-the-art quantum experiments with laser cooled quantum gases, which allow for $\sim0.5$~m  separation between the superposed paths maintained for $\sim$1~sec~\cite{Kovachy2015}, are still not sufficient to observe reduced visibility due to the gravitational time dilation.

Whereas testing the effects of the gravitational time dilation on interfering quantum clocks will remain challenging in the near future, testing the special relativistic effects appears within the reach of the already existing technology. In the following we describe a relativistic clock interferometer implemented with a single trapped electron.

\section*{Single electron quantum cyclotron}
 A setup that allows the trapping of a single electron is  a Penning trap. There a large permanent magnetic field of $B=5$~T is applied to achieve radial confinement and a DC quadrupole electric potential provides axial trapping~\cite{Brown1986}. The radial cyclotron motion of an electron proceeds with orbital frequency $\omega_{c} = eB/m = 2\pi\times140$~GHz, where $m$ is the mass of the electron, and it is cooled via synchrotron radiation. At temperatures below 100~mK, the cyclotron motion resides near its quantum-mechanical ground state (lowest Landau level) with probability $P > 0.999$ until a resonant excitation is applied, so that it must be treated quantum mechanically~\cite{Peil1999}. The axial motion of the trapped electron has a characteristic frequency $\omega_z /2 \pi= 100$~MHz frequency and it is cooled and detected by using a resonant LC-circuit~\cite{Brown1986, Durso2003}. In existing electron trapping experiments (as well as in proton and ion experiments) cyclotron and spin states are measured by using the magnetic bottle technique. The parabolic profile of the field along the trap axis couples axial and cyclotron motions. The frequency of the axial oscillations depends linearly on the energy of the cyclotron and spin states and allows for their quantum non-demolition (QND) measurement or quantum-jump spectroscopy. By using this QND technique it was possible to attain remarkable accuracy of few parts in 10$^{-13}$ in the determination of the electron magnetic moment~\cite{Odom2006,Hanneke2008}.

A single electron in a Penning trap can directly be described by the Dirac equation in an external electromagnetic field. A detailed derivation of the resulting dynamics is given in ref.~\cite{Brown1986}, and to lowest order in relativistic corrections, the Hamiltonian of the trapped electron reads
\begin{equation}\label{Relativistic_H}
H_{rel}=\frac{\vec{\Pi}^2}{2m}\left(1-\frac{\vec{\Pi}^2}{4mc^2}\right)+eV+\vec{\mu}\vec{B}\left(1-\frac{\vec{\Pi}^2}{2m^2c^2}\right), 	
\end{equation}
where $\vec{\Pi}$ is the kinetic momentum and the first term describes the relativistic energy of the cyclotron degree of freedom, $eV$ is the scalar potential energy, and the last term describes the magnetic energy with a relativistic correction. The magnetic moment of the electron is given by $\vec{\mu}=-\frac{\textrm{g}e\hbar}{4m}\vec{\sigma}$ with $\vec{\sigma}$ the Pauli matrices and g is the electronic g-factor. The Hamiltonian $H_{rel}$ and a relativistic-clock Hamiltonian derived in refs.~\cite{Zych2011, Pikovski2015} are the same at this order (neglecting gravitational effects in the latter). The term $\vec{\mu}\vec{B}\left(1-\frac{\vec{\Pi}^2}{2m^2c^2}\right)$ is of key importance for the present work as it describes the clock of the electron, i.e. spin precession, and its special relativistic time dilation.

\begin{figure}[ht!]
\includegraphics[width=1\columnwidth]{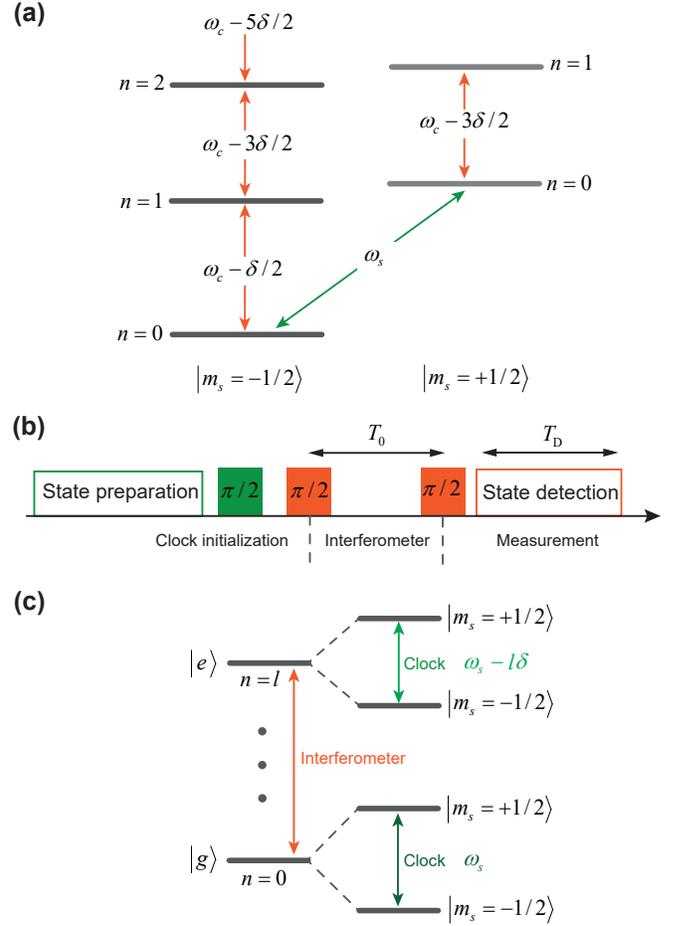}
\caption{(Color online) \textbf{(a)} Energy level structure of a trapped electron in a Penning trap. Here, $\omega_c=eB/m$ is the non-relativistic cyclotron frequency, where $e, m, B$ are the electron charge, mass, and the magnitude of the magnetic field, respectively;  $\delta= \omega_c \frac{\hbar \omega_c}{m c^2}$ describes the relativistic frequency shift and $\omega_s=\omega_c(\textrm{g/2})-\delta/2$ is the spin frequency for $n=0$, g is the electronic g-factor. \textbf{(b)} Time sequence for the clock interference experiment. Cyclotron states with $n=0$ and $n=l$ are chosen for the Ramsey interferometer. \textbf{(c)} Spin states of the trapped electron will serve as a clock. Due to time dilation the clock frequency depends on the cyclotron quantum number $n$. The beating between such relativistically shifted spin frequencies for the two superposed cyclotron states, will result in the modulation of the amplitude of the interference pattern (e.g. obtained by changing $T_0$).}\label{Spectrum}
\end{figure}

The magnitude as well as the qualitative understanding of the relativistic effects can be conveniently characterised from the spectrum of $H_{rel}$, for which we give a simple justification below. A complete derivation thereof can be found ref.~\cite{Brown1986}.  Up to the constant potential $eV$, in the non-relativistic limit the spectrum of the trapped electron is given by the sum of the energies of the cyclotron states $\frac{\Pi^2}{2m}=\hbar \omega_c(n+\frac{1}{2})$, where $n=0,1,...$ and of the spin $\vec{\mu}\vec{B}=\hbar\omega_cm_s$, with $m_s=\pm\frac{1}{2}$. Thus, one  immediately finds that with relativistic corrections the spectrum reads $\hbar\omega_c(n+\frac{1}{2})[1-\frac{\hbar\omega_c}{2mc^2}(n+\frac{1}{2})]+\hbar\omega_cm_s[1-\frac{\hbar\omega_c}{mc^2}(n+\frac{1}{2})]$, and consists of two different, slightly anharmonic ladders of cyclotron states for the two spin projections, see Fig.~\ref{Spectrum}(a). 

The above shows that the frequency of the cyclotron transitions, between levels $n$ and $n+1$, experiences level- and spin-dependent relativistic shifts: it reads $\omega_c+\delta_c$, where  $\delta_c= -\omega_c \frac{\hbar \omega_c}{m c^2}(n+1+m_s)$. The spin transition frequency,  on the other hand, now reads $\omega_c+\delta_s$ where $\delta_s=-\omega_c \frac{\hbar \omega_c}{m c^2}(n+\frac{1}{2})$ --  it is shifted depending on the cyclotron quantum number $n$. In other words, the spin transition frequency is redshifted by a factor $1-\frac{\hbar \omega_c}{m c^2}(n+\frac{1}{2})$. In a full analogy to the classical case a ``faster moving'' clock (here: electron with a higher energy, larger $n$) appears to tick slower than an identical clock ``moving slower'' (here: lower $n$). The frequency shift $\delta_s$  is a direct expression of the special relativistic time dilation of the electron spin precession. The magnitude of all the above shifts is described by a single quantity
\begin{equation}\label{delta_l}
\delta= \omega_c \frac{\hbar \omega_c}{m c^2}.
\end{equation}

At a magnetic field of $5$~T the resulting relativistic spin frequency shift for the ground cyclotron state $\delta_s\sim 2\pi\times150$~Hz corresponds to the redshift factor obtained for a classical linear motion with a velocity $v_{0}\sim 10$~km/s and exceeds the radiation emission rate in free space
\begin{equation}\label{gamma0}
T_1^{-1}=\gamma_0 \approx \frac{1}{4\pi\epsilon_0}\frac{4e^2 \omega_c^2}{3mc^3} \simeq 2\pi\times2~\textrm{Hz}.
\end{equation}
Therefore, a single trapped electron offers an ideal platform for realization of a relativistic clock interference experiment: the clock could be implemented in a superposition of the electron spin projections in the direction of the trap field, and the path superposition  -- in a superposition of cyclotron states with different quantum numbers, e.g.~$n=0$ and $n=l$, see Fig.~\ref{Spectrum}(b,c). As the clock frequency is time dilated by a different amount $\delta_s$ for the different cyclotron orbits,  such an interference experiment would indeed probe a clock that ticks in a superposition at different rates due to time dilation.

The interferometer sequence requires coherent control of cyclotron and spin states of the electron which has not been demonstrated yet. Hanneke et al.\ attempted to encode the qubit in cyclotron states $n = 0$ and $n = 1$ with the aim to decrease the wait-time in quantum jump spectroscopy~\cite{HannekePhD}. All efforts at seeing Rabi oscillations in these experiments has been thwarted by two conflicting timescales. The first one is the Rabi frequency  $\Omega^c_R$ which should not exceed the anharmonicity of the electronic states, i.e. should satisfy  $\Omega^c_R <\delta$, if one wants to address a single cyclotron transition. The second time scale is related with the magnetic bottle which causes the cyclotron line broadening by $\sim 300$~Hz~\cite{Hanneke2011}, which exceeds both $\delta$ and $\Omega_R^c$. Thus, to establish coherent control of cyclotron and spin states of a trapped electron, narrower lines or larger relativistic shift are necessary. In the sections below we address both these issues.

\section*{Single electron relativistic clock interferometer}
Here we describe experimental techniques for realization of the relativistic clock interferometer with a single electron cooled to the cyclotron ground state of the Penning trap. As discussed in the section above, we consider clock implemented in the spin degree of freedom of the electron  (explained in detail in the following sections) and the clock's path described by the cyclotron state.

The clock state preparation will be followed by the Ramsey interference between cyclotron states. The maximal evolution time of the clock in the interferometer $T_0$ (time between two $\pi/2$ pulses) should not exceed the cyclotron $T_2^c$ and spin $T_2^s$ coherence times. The phase coherence times of the cyclotron and spin degrees of freedom for trapped electron have not yet been measured. The spin flip time is a few years and can serve as the upper bound for $T_2^s$. The cyclotron coherence time $T_2^c$ gives a much more stringent constraint on the evolution time due to the decay of the cyclotron state at the rate $\gamma_0$, see Eq.~\ref{gamma0}, so for higher cyclotron orbit $T_2^c \leq 2(\gamma_0 l)^{-1}$. Since $T_2^c$ is not yet known, we assume that $T_2^c \sim (\gamma_0 l)^{-1}$. Under such assumption the maximum evolution time in the interferometer is $T_0 = (\gamma_0 l)^{-1}$. The maximal proper time difference attainable in such an interferometer is
\begin{equation}\label{proper time}
\Delta \tau = T_0{\frac{\Delta E}{mc^2}},
\end{equation}
where $\Delta E = \hbar \omega_c l$ is the energy difference between {the} cyclotron states. Therefore, we expect to observe $N_m$ disappearing/revival periods of Ramsey interference induced by {the special} relativistic time dilation:
\begin{equation}\label{Number of fringes}
N_m = \Delta \tau/T_{clock}=\omega_c \Delta \tau /\pi = \frac{3}{8\pi\alpha}\approx 16,
\end{equation}
where $\alpha$ is the fine structure constant. It is interesting to note that $N_m$ does not depend either on cyclotron frequency nor on the quantum number of the upper cyclotron state. Therefore, any convenient value of the cyclotron frequency $\omega_c$ and quantum number $l$ can be chosen for the relativistic clock interferometer. Even for order of magnitude smaller coherence times the proposed setup will still allow measurement of a few oscillations of the visibility of Ramsey interference.

\section*{Transmission spectroscopy of geonium atom}
In this section we describe the proposal for detection of a single electron quantum state without the use of a magnetic bottle. For the detection of cyclotron transitions we apply a method initially developed for the detection and manipulation of superconducting qubits strongly coupled to a 1D transmission line~\cite{Astafiev2010}. There, a strong interaction between an artificial atom and radiation field confined in 1D space results in high extinction of an excitation signal. By applying pulsed excitation it was also possible to observe coherent and incoherent dynamics of an artificial atom~\cite{Abdumalikov2011}.

\begin{figure}[ht!]
\includegraphics[width=1\columnwidth]{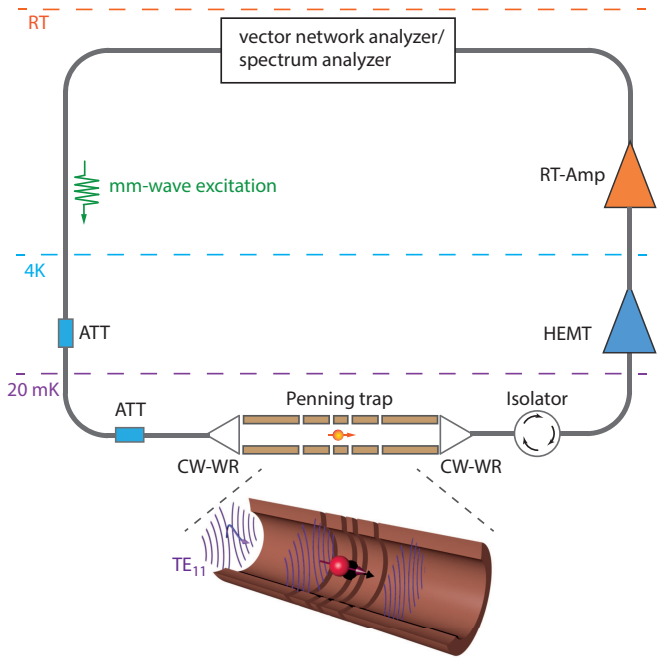}
\caption{(Color online) Experimental setup. Single electron is trapped in an open-endcap Penning trap. The magnetic field is directed along the trap axis. For the detection of a geonium atom we propose using low-noise transmission spectroscopy at the excitation power corresponding to a few mm-wave photons. The excitation signal from a vector-network analyzer is cold attenuated and is guided into the Penning trap with the help of rectangular waveguide-to-circular waveguide converters (WR-CW). The cylindrical electrodes of the trap of the proper dimensions will play a role of TE$_{11}$ waveguide for the excitation and the emitted signal. The transmitted and/or the emitted signals will pass through a low-noise detection setup formed by mm-wave isolator(s) thermally anchored to mixing chamber of a dilution fridge and will be amplified by cryogenic HEMT amplifier (HEMT) and room temperature (RT-Amp) amplifiers. A spectrum analyzer can be used for the detection of cyclotron fluorescence.}\label{Setup}
\end{figure}

The sketch of the proposed setup is shown in Fig.~\ref{Setup}. A single electron will be confined and cooled in an open-endcap cylindrical Penning trap~\cite{Gabrielse1989}. The diameter of the trap electrodes can be made such that it matches the dimensions of a cylindrical waveguide for TE$_{11}$ mode for a signal at $\omega_c$. In order to minimize a reflection of a mm-wave signal from the joints between the trap electrodes, the gap size between them, $z_g$, should satisfy: $z_g\ll \lambda_c$; where $\lambda_c$ is the wavelength of the signal at cyclotron frequency $\omega_c$. In the typical Penning trap $z_g\approx0.1$~mm which is 1-3\%$\lambda_c$, therefore, the return loss for each pair of the trap electrodes is expected to be $|S_{11}|< -15$~dB~\cite{Pucci2012}. Thus, the trap apparatus forms an effective 1D transmission line for the synchrotron radiation emitted by a artificial atom, which is also called geonium~\cite{Brown1986}.

The geonium can be detected in a mm-wave transmission measurement with the help of a vector network analyzer (VNA), where the strong confinement of the radiation within the emission wavelength $\lambda_c$ results in full extinction of the probing signal. To avoid saturation of the dipole transition between the cyclotron states $n=0$ and $n=1$ it should be probed at a low excitation power $ P_f$ corresponding to the photon number between $1$ and $10$ per emission cycle, that is $P_f\simeq\mathrm{photon\;number}\times\hbar \omega_c /T_1$ between $10^{-21}~W$ and $10^{-20}$~W, where $T_1$ is the emission time of a cyclotron quantum. For that purpose the coherent excitation signal passes through cold attenuators. The signals below 50~GHz can be delivered by using coaxial cables, whereas for the signals above 50~GHz teflon stripe of the proper dimension can be used. The incoming signal will be coupled to a cylindrical waveguide with transition section (CW-WR). After interaction with geonium atom, the transmitted signal is initially amplified with a cryogenic HEMT amplifier, separated from the trap apparatus with a mm-wave isolator. One or two additional room temperature amplifiers (RT-Amps) will bring the signal above the noise level of the VNA. Since the power of the resonantly emitted mm-wave photons is smaller than the thermal noise of the amplifier, in order to detect geonium extinction, the measured signal shall be averaged out. In that case the minimal required detection time per single frequency point is estimated to be
\begin{equation}\label{VNA Detection Time}
T_{det}\sim\frac{1}{\Delta f}\left(\frac{k_B T_n \Delta f}{P_f}\right )^{m},
\end{equation}
where $\Delta f$ stands for the resolution bandwidth, $T_n$ is the temperature of the HEMT amplifier, $m=1$ when the signal detected with VNA and $m=2$ when the signal detected with the spectrum analyzer (SA). For the VNA detection of geonium atom at $\omega_c=2\pi\times140$~GHz the measured extinction spectrum is assumed to consist of 5-10 points. For that purpose the resolution bandwidth of VNA is set to $\Delta=1$~Hz. If the noise temperature of mm-wave HEMT amplifier is $T_n \sim 40$~K then $T_{det}\simeq 100$~sec for the full spectrum. The long detection time imposes certain constraints on the stability of the experiment. For example, during the detection of the spectrum the magnetic field should be stable within $10^{-10}$, which is technically feasible, s.f.~\cite{Hanneke2011} and references therein.

The above described transmission mm-wave spectroscopy of a single electron cyclotron allows one to circumvent complex parameter estimations and fits in the measuring of the cyclotron frequency. Therefore it might result in even higher precision in measurements of fundamental constants~\cite{Odom2006,Gabrielse2006}.

Since the emission rate of synchrotron radiation $\gamma_0\propto \omega_c^2$ and the ratio $k_B T_n/\hbar \omega_c$ is usually constant for quite a broad frequency range for the most of cryogenic HEMTs, the detection time $T_{det} \propto 1/\omega_c^{2m}$. To attain the shorter detection times it is desirable to work at higher frequencies.  However, at the present time, cryogenic HEMT amplifiers are commercially available for the frequency range $28-42$~GHz and with $T_n\sim15$~K. Cryogenic HEMTs amplifiers for the W-band (75-110~GHz) with $T_n\sim30$~K are technically feasible and will be commercially available in the near future~\cite{Samoska2011}. Below we will also consider two possible realizations of coherent control for high frequency of 110~GHz and for lower frequency of 40~GHz geoniums.

\section*{Coherent control of cyclotron and spin states of high-frequency geonium}

\begin{figure}[ht!]
\includegraphics[width=1\columnwidth]{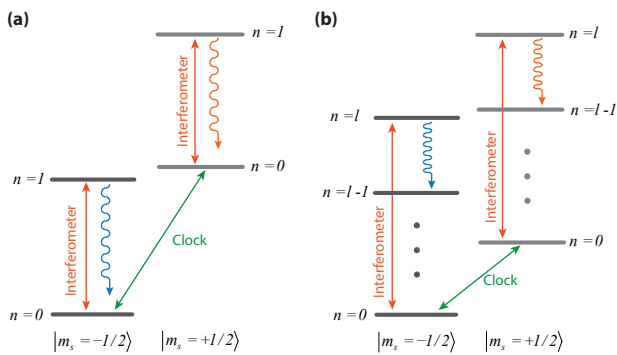}
\caption{(Color online) Realization of the single electron clock interferometer. \textbf{(a)} At $\omega_c/2\pi=110$~GHz a two level system can be realized between states $n=0$ and $n=1$. The read-out of cyclotron and spin states is made by measuring the fluorescence from the excited state $n=1$. \textbf{(b)} At $\omega_c/2\pi=40$~GHz the qubit will be realized between states $n=0$ and $n=l$, where $l>1$. For the cyclotron state read-out the fluorescence signal from the excited state $n=l$ to $n=l-1$ is measured. The read-out of spin states is performed via extinction measurements on cyclotron $0\leftrightarrow1$.}
	\label{Qubit}
\end{figure}

For cyclotron frequency $\omega_c/2\pi \simeq 110$~GHz the effective two-level system for the interferometer can be encoded between the states $n=0$ and $n=1$, see Fig.~\ref{Qubit}(a). The anharmonicity, i.e. the non-linearity of the cyclotron harmonic ladder, between $0\leftrightarrow 1$ and $1\leftrightarrow 2$ transitions is equal to $\delta/2\pi\simeq 115$~Hz, which in turn limits any coherent drive between the states to Rabi frequencies of $\Omega_R^c\leq\delta$ in order to stay within the two-level subsystem. The coherent control of cyclotron states will be established by applying a transient pulse of a certain duration followed by the measurement of the mm-wave fluorescence from $n=1$ state. To attain $\Omega_R^c/2\pi \sim 100$ Hz the required power level of the pulse delivered into the Penning trap is estimated to be
\begin{equation}\label{Power_to_manipulate_cyclotron}
P_{cyc}\sim \frac{c\epsilon_0}{2}\left(\frac{\hbar \Omega_R^c}{d_{01}}\right)^2 \lambda_c^2 \sim 10~\textrm{aW}.	
\end{equation}
Here, the $d_{01}=e \sqrt{\hbar/2 m \omega_c}\simeq200~ea_0$ is the dipole moment for the cyclotron transition $0\leftrightarrow 1$, where $a_0$ is the Bohr radius. After coherent manipulation, the state read-out is performed by measuring the power spectrum of fluorescence emitted from the excited state~\cite{Gerhardt2007}. Such incoherent emission can be measured with the help of the spectrum analyzer. The excited states of the geonium with different spin projections, i.e.\ the states $|1, -1/2\rangle$ and $|1, +1/2\rangle$, can be discriminated by measuring the frequency of the fluorescence signal, because they are shifted by $\delta$ with respect to each other.

The coherent manipulation of the spin is performed in a similar way. For that purpose the transient pulse resonant with the spin-flip transition is applied to the waveguide trap. The power required for the coherent spin manipulation with Rabi-frequency $\Omega_R^{s}$ is
\begin{equation}\label{Power_to_manipulate_spin}
P_{s}\sim \frac{c}{2\mu_0}\left(\frac{\hbar \Omega_R^{s}}{2\mu_B}\right)^2 \lambda_c^2,	
\end{equation}
where $\mu_B$ is the Bohr magneton. The manipulation sequence is followed by the state read-out with the help of VNA. It probes the extinction around qubit transitions at $\omega_c-\delta/2$ and $\omega_c-3\delta/2$ at a single point with resolution bandwidth $\Delta=\gamma_0$. In order to acquire statistics the sequence is repeated and the ratio of the extinction magnitudes will yield the probability of the spin-flip. Another spin read-out method is to apply a $\pi$-pulse on the cyclotron transition and measure the fluorescence.

The implementation of the clock interferometer relies on manipulation of the spin and the cyclotron states at the time scale much shorter than $T_1$. The manipulation time of the spin state (clock) is mainly limited by the input power of the signal. To attain $\Omega_R^{s} \sim 1$~kHz the power level of $\sim 1~\mu$W is required. The mm-wave signal for spin can be delivered to the trap by using the teflon stripe which fits the dimensions of WR-10 waveguide. Transmission losses of the signal in the 1~m teflon waveguide of about 3~dB are expected, where most of losses occur in the coupling of the stripe to WR waveguide. The typical output power of the mm-wave frequency extender for W-band is about 10~mW. By taking into account $\sim30$~dB cold attenuation, the estimated maximum power level of the mm-wave signal at the input of the trap is $\sim 10$~$\mu$W. Thus, Rabi frequencies for the spin drive of up to 3~kHz are technically attainable.

The relatively fast manipulation of the cyclotron states is more challenging. On one hand, in order to avoid the leakage of the population to the next cyclotron level during sufficiently strong driving, the length of the resonant $\pi/2$ pulse $t_{\pi/2} > 1/\delta \simeq 10$~ms. On the other hand, the time required to manipulate the cyclotron states should be much less than the synchrotron radiation time, so $t_{\pi/2} \ll T_1 \sim 100$~ms. In principle, all the above mentioned conditions for the pulse length are fulfilled for the proposed experiment with encoding the cyclotron interferometer between $n=0$ and $n=1$ at the resonance frequency of 110~GHz.

\section*{Detection and coherent control of low-frequency geonium}
In this section we consider implementation of the clock interferometer between the ground and higher lying cyclotron states $n=2,3,4,..$. Here, for the illustration of the proposed method we consider the geonium atom with $\omega_c = 2\pi\times40$~GHz. The interferometer will be realized by using the $0 \leftrightarrow 5$ transition at 200~GHz. This method will take an advantage of the increased anharmonicity with increasing energy. Another advantage of using higher lying cyclotron levels is that the decay time of the excited state is $l$ times smaller than the decay time of the interferometer implemented at the same energy difference but using $0\leftrightarrow 1$ transition.

The detection of geonium at 40~GHz frequency is more challenging due to smaller fluorescence power $P_{f}\sim 10^{-22}$~W in comparison to 110~GHz and hence longer integration time $T_{det}$. The detection time of the extinction spectrum of geonium atom with the help of VNA at $\Delta f \simeq 0.1$~Hz is estimated by using Eq.~\ref{VNA Detection Time} and is $T_{det}\sim1000$~sec. During this time the cyclotron frequency shall be stable within the detection bandwidth yielding thus the upper limit of the fluctuation level of the magnetic field of $10^{-11}$, which is more strict than in the case of 110~GHz geonium but still attainable. After the detection of geonium atom with VNA and determination of transition frequencies, one may establish coherent control of both the cyclotron and spin states.

The read-out of the spin state is performed in the same way as for the previously described high-frequency geonium. However, the cyclotron read-out requires an alternative method, see Fig.~\ref{Qubit}(b). The transient mm-wave pulse resonant with cyclotron transition $0\leftrightarrow l$ will prepare one of the excited cyclotron states $n=l$, and the power of synchrotron radiation emitted at $l\leftrightarrow l-1$ transition is measured. By repeating the sequence many times, the emitted power will yield the probability of the excited cyclotron state and will be detected by using the spectrum analyzer. After the state detection is finished, the electron should be prepared in its ground state by cooling via synchrotron radiation. The time required for ground state preparation is estimated to be $T_p>2~T_1 \ln (l) \simeq 3$~sec. The manipulation sequence is repeated and fluorescence spectrum is measured again yielding the required probability.

One of the crucial challenges in the implementation of the interferometer between higher cyclotron states is the magnitude of the dipole moment for the direct transition between $n=0$ and $n\ge2$. The extent to which the relatively small relativistic corrections will allow for such a direct transition is not yet explored. Here, instead, we propose using optimal control techniques to prepare a higher lying Fock state of the weakly anharmonic oscillator~\cite{Motzoi2009,Gambetta2011} without populating other levels. This and similar techniques have recently been used to control state leakage for superconducting based qubits~\cite{Schutjens2013,Chen2016}, which are also weakly anharmonic oscillators. Although the anharmonicity of the geonium states is several orders of magnitude weaker than in the present experimental demonstrations, the fidelity of the coherent population transfer required is also orders of magnitude less stringent than that required for superconducting qubit gates. Although not such a necessary requirement, such techniques can also be used to improve the fidelity of pulses in the high frequency ($n=0$ to $n=1$ transition) geonium realisation.

\section*{Entanglement of two electrons}
\begin{figure}[ht!]
\includegraphics[width=1\columnwidth]{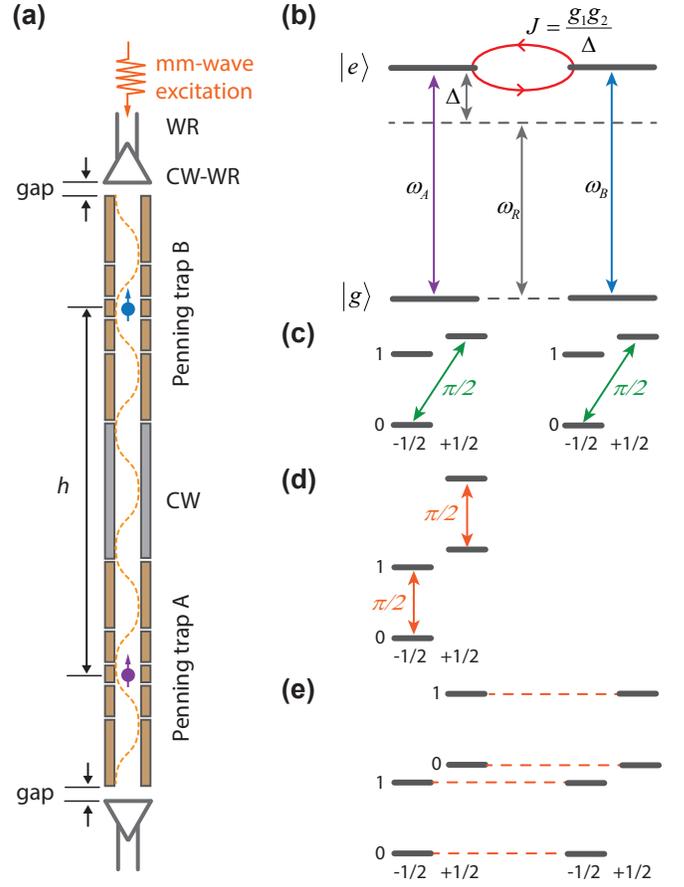}
\caption{(Color online) Entanglement of two electrons.\textbf{(a)} Two electrons will be trapped in separated Penning traps. The traps are connected with narrow-gap circular waveguide (WR). The mm-wave cavity will be formed by trap apparatus decoupled from the waveguide with the variable gap, which is $\sim \lambda_c$. \textbf{(b)} Spectroscopic scheme for the dispersive geonium-geonium coupling. \textbf{(c)} Sequences for generation of entanglement between two geonium atoms. Strong $\pi/2$-pulse prepares the clock states of twin A and twin B. \textbf{(d)} A superposition of cyclotron states of one of the twins will be prepared by applying a $\pi/2$-pulse. \textbf{(e)} Twin A and twin B are tuned in resonance with each other for time $T_g = \pi/2J$ necessary to generate the entanglement. }
	\label{Entanglement_Setup}
\end{figure}
Consider now a \textit{pair} of clocks on two paths with different proper times, in a state where the paths followed by the clocks are entangled. The proper times elapsed for the clocks should as a result become entangled as well and one could in principle violate the Bell inequality  making measurements on the electrons' spins. A violation of the inequality would show that the time elapsed for a physical system cannot always be described by a local (only referring to the given clock) parameter. Scenarios where the evolution time of a system is not describable by a classical parameter, e.g.~becomes uncertain, are discussed in the context of quantum gravity and are often considered unphysical or assumed to be necessary causing decoherence, see ref.~\cite{Diosi2005timedecoh}. Realization of the Bell experiment for the proper time would not only show that such situations are experimentally accessible, but could also help building the intuition required for their description in a more general contexts than the experiment itself.

In this section we describe such an experiment with a pair of electron clocks entangled in their cyclotron degrees of freedom. The clocks will be implemented in the spin superposition states $|c\rangle_{A,B}=\left( |-1/2\rangle_{A,B}+|1/2\rangle_{A,B}\right)/\sqrt{2}$, where $A$ ($B$) label the two electrons. The paths will be implemented in the cyclotron states $|0\rangle_{A,B}$ and $|1\rangle_{A,B}$. Thus, the entangled state of the clocks to be implemented reads
\begin{equation}\label{Entangled_state}
\Psi_{A,B}\propto|c, 0\rangle_A|c, 1\rangle_B+i|c, 1\rangle_A|c, 0\rangle_B.
\end{equation}

To engineer the state $\Psi_{A,B}$ we plan to interface two geonium atoms coupled to the same cavity. Here, we will consider entanglement generation protocol for high-field geoniums with the cyclotron frequency of 110~GHz. The sketch of the proposed setup is shown in Fig.~\ref{Entanglement_Setup}(a). It consists of two waveguide Penning traps interconnected by a circular waveguide section. Traps are equipped with additional small coils for the fast tuning $\sim 1$~mT/ms of the trapped electrons. Separation distance can be chosen $h\sim0.1-1$~m and it is mainly limited by the dimension of the dilution fridge and a magnet, which should create a uniform and highly stable field in Penning traps. Two gap sections $\sim \lambda_c$ at the entry and at the exit of the waveguide or small pinholes can be used to form a resonator for mm-waves. The present state-of-the-art suggest the quality factor to be $Q_c\sim10^3-10^4$~\cite{Hanneke2011}. In the following we assume that the quality factor is $Q_c\sim10^4$ yielding cavity decay rate of $\kappa_c/2\pi\sim10$~MHz.

The dipole-dipole interaction between cyclotron states via the effective 1D transmission line is too weak to generate entanglement $g_{dd} \sim \gamma_0/2$. Therefore, in order to entangle remote geonium atoms we propose to use the interaction mediated by the exchange of a virtual photon confined in the cavity~\cite{Majer2007}. When two atoms are tuned away by $\Delta$ from the resonance frequency of the cavity $\omega_r$ but in resonance with each other they interact via virtual exchange of mm-wave photons, see Fig.~\ref{Entanglement_Setup}(b). The interaction coupling strength is $J=g_1 g_2 /\Delta$, where $g_{1,2}/2\pi\sim0.1$~MHz are estimated coupling strengths of the cyclotron states to the cavity of length of 30~cm. For the detuning of $\Delta/2\pi\sim50$~MHz the coupling strength is $J/ 2\pi\sim200$~Hz, which is much larger than the decay rate of the cyclotron states $\gamma_0/2\pi\simeq1.5$~Hz.

The entanglement generation sequence is presented in Fig.~\ref{Entanglement_Setup}(c-e). Two electrons are trapped in their individual traps and cooled down to their ground state $|n_c=0,-1/2\rangle$. The clocks are prepared by applying a strong $\pi/2$-pulse resonant with spin-flip transitions of the twins. Then twin B is detuned by more than $\kappa_c$. A $\pi/2$-pulse is applied in resonance with cyclotron $0\leftrightarrow1$ transitions, thus creating a superposition between clock and cyclotron states for the twin A. After that, twin B is tuned in resonance with the twin A and after time $\pi/2J$ their cyclotron and spin states will be entangled yielding the required state $\Psi_{A,B}$ .

The proposed experimental techniques can be further developed for trapping of a positron, i.e. an ``anticlock'', and for entangling an electron and a positron. A single electron and a single positron can be trapped in individual traps, analogously as sketched in Fig.~\ref{Entanglement_Setup}(a) for a pair of electrons. Their degrees of freedom can be entangled by using the above described procedure. Experiments with ``anticlocks'', or with an entangled clock-anticlock pair, will pave the way towards conceptually new tests of time-reversal, which is an interesting study for future research.

\section*{Conclusion}
Since the first demonstration of the single trapped electron in 1973, experiments with this genuine quantum system have been driven by the idea of precision tests of Quantum Electrodynamics (QED)~\cite{Wineland1973}. For example, in such experiments the magnetic moment of the electron has been measured with unprecedented precision of $\sim10^{-13}$~\cite{Hanneke2008}. In this article, we showed that a single trapped electron can also be used for new tests of the notion of time in a regime where it requires relativistic as well as quantum description. For that purpose, we described a relativistic clock interferometer based on a single electron quantum cyclotron. The clock interferometer requires coherent control of a quantum electron cyclotron which has not been demonstrated yet. However, here we proposed a novel detection and coherent manipulation technique based on coupling of the electron cyclotron to an effective 1D transmission line, which represents a feasible experimental route in this direction.
The described single electron relativistic clock interferometer will allow for the first experimental test of the effects of proper time on quantum coherence. In particular, it will answer the question whether proper time of a clock can be prepared in a quantum superposition~\cite{Zych2011}. Moreover, our proposed setup with a pair of entangled electrons will allow for a Bell tests for the proper time of entangled clocks~\cite{Zych2012}. Relativistic interfering clocks can also verify theories where time is considered as a new quantum degree of freedom~\cite{Kudaka1999}. Finally, realization of our experiments will yield evidence on the role of time dilation as a decoherence mechanism~\cite{Pikovski2015}. Implementation of the relativistic clock interferometer will provide first experimental insights into the still open, fascinating questions about the physics of time~\cite{Smolin2014}.

We thank S. Probst, M. Henkel and G. Morigi for the stimulating discussion and critical reading of the manuscript.
This work was supported by the Australian Research Council Centre of Excellence for Engineered Quantum Systems grant number CE110001013 and the University of Queensland through UQ Fellowships grant number 2016000089.

\bibliographystyle{apsrev}
\bibliography{electron}

\end{document}